%% file: main.tex
\documentclass[10pt,twocolumn,letterpaper]{article}

\usepackage{iccv}              


\definecolor{iccvblue}{rgb}{0.21,0.49,0.74}
\usepackage[pagebackref,breaklinks,colorlinks,allcolors=iccvblue]{hyperref}

\def\paperID{12427} 
\def\confName{ICCV}
\def\confYear{2025}

\title{ARAP-GS: Drag-driven As-Rigid-As-Possible 3D Gaussian Splatting Editing with Diffusion Prior}

\author{Xiao Han$^{1,}$\thanks{Equal contributions to this work.}\quad
\and
Runze Tian$^{1,}$\footnotemark[1]\quad
\and
Yifei Tong$^{1,}$\footnotemark[1]\quad
\and
Fenggen Yu$^{2}$\quad
\and
Dingyao Liu$^{1}$\quad
\and
Yan Zhang$^{1}$\quad
\and
$^{1}$Nanjing University\quad 
$^{2}$Simon Fraser University\\
{\tt\small hanxiao@smail.nju.edu.cn}
}

\begin{document}
\twocolumn[{%
\renewcommand
\twocolumn[1][]{#1}%
\maketitle

\centering
\includegraphics[width=1\linewidth]{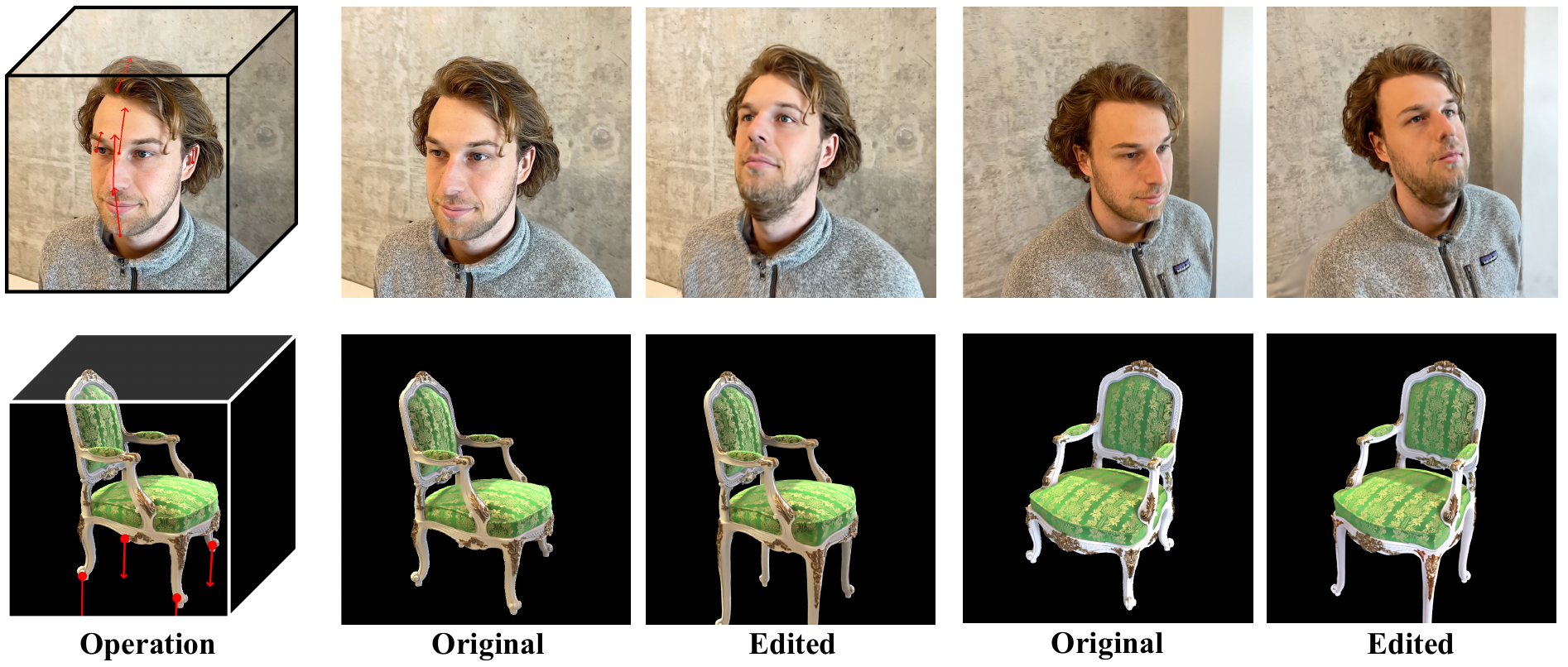}
\captionof{figure}{
    \textbf{\textit{Results of ARAP-GS.}} Given a set of handle points and their deformations, ARAP-GS can efficiently achieve drag-driven 3DGS editing. Our method deforms the geometry of the 3DGS scene through rotation (above) or stretching (below) while preserving the original appearance and multi-view consistency. The first column illustrates the dragging operation, with the red points indicating the handle points and the arrows indicating the dragging directions. "Original" and "Edited" denote the rendering results before and after editing.
\vspace{1em}
}
\label{fig:teaser}
}]
\begingroup
\renewcommand\thefootnote{}\footnote{*Equal contributions to this work.}%
\addtocounter{footnote}{-1}
\endgroup 
\input{sec/0_abstract} 
\input{sec/1_intro}
\input{sec/2_relatedwork}
\input{sec/3_method}

\input{sec/4_experiments}
\input{sec/5_conclusion}
{
    \small
    \bibliographystyle{ieeenat_fullname}
    \bibliography{main}
}


\end{document}

%% file: sec/0_abstract.tex
\begin{abstract}
\label{abstract}
Drag-driven editing has become popular among designers for its ability to modify complex geometric structures through simple and intuitive manipulation, allowing users to adjust and reshape content with minimal technical skills. This drag operation has been incorporated into numerous methods to facilitate the editing of 2D images and 3D meshes in design. However, few studies have explored drag-driven editing for the widely-used 3D Gaussian Splatting (3DGS) representation, as deforming 3DGS while preserving shape coherence and visual continuity remains challenging. In this paper, we introduce ARAP-GS, a drag-driven 3DGS editing framework based on As-Rigid-As-Possible (ARAP) deformation. Unlike previous 3DGS editing methods, we are the first to apply ARAP deformation directly to 3D Gaussians, enabling flexible, drag-driven geometric transformations. To preserve the scene appearance after deformation, we incorporate an advanced diffusion prior for image super-resolution into our iterative optimization process. This approach enhances visual quality while maintaining multi-view consistency in the edited results. Experiments show that ARAP-GS outperforms current methods across diverse 3D scenes, demonstrating its effectiveness and superiority for drag-driven 3DGS editing. Additionally, our method is highly efficient, requiring only 10 to 20 minutes to edit a scene on a single RTX 3090 GPU.
\end{abstract}

%% file: sec/1_intro.tex
\section{Introduction}
\label{sec:intro}
Recently, 3D Gaussian Splatting (3DGS)~\cite{kerbl20233d} has gained significant attention in 3D vision, enabling breakthroughs in novel view synthesis, image-based 3D reconstruction, and other vision tasks due to its explicit representation and real-time rendering. As the technology advances, it has been extended to various tasks, including dynamic scene modeling~\cite{lu20243d, lee2024fully}, 3D scene understanding~\cite{bai2024360, shi2023gir}, and 3D scene editing~\cite{chen2024gaussianeditor, palandra2024gsedit}.

Existing 3D scene editing methods aim to provide users with a precise and intuitive editing framework, allowing them the freedom to adjust every detail of a scene. This includes not only texture editing but also flexible geometric deformation. While 3D scene editing has been extensively studied on 3D meshes, few methods can be directly applied to the 3DGS representation, as deforming 3DGS while preserving shape coherence and visual continuity remains challenging.
Recently, researchers have explored various editing operations for 3DGS, such as text-driven 3DGS editing~\cite{wu2024gaussctrl, zhuang2024tip, wang2025view}, monocular-video-driven 3DGS editing~\cite{cai2024dynasurfgs, huang2024sc}, and cage-driven 3DGS editing~\cite{huang2024gsdeformer}. Unlike these works, we aim to develop an efficient drag-driven 3DGS editing approach that enables flexible geometric transformations for 3DGS, as shown in \cref{fig:teaser}.

Drag-driven editing is popular among designers for its intuitive, user-friendly approach to modifying complex geometric structures, allowing users to easily adjust and reshape content with minimal technical expertise, and has been widely used in 3D mesh editing~\cite{sorkine2007rigid,chao2010simple,levi2014smooth, chen2017rigidity}. With the rapid development of 2D generative models~\cite{goodfellow2014generative, rombach2022high}, several drag-driven image editing methods have also emerged, such as DragGAN~\cite{pan2023drag} and DragDiffusion~\cite{shi2024dragdiffusion}. These methods propose a revolutionary interactive image manipulation framework that enables users to "drag" any point to a target position, providing precise control over pose, shape, expression, and layout. However, synthesizing multi-view images to maintain geometric consistency is suboptimal, as directly editing Gaussian attributes leverages 3DGS’s explicit representation more effectively.

In this paper, we propose ARAP-GS, an intuitive drag-driven 3DGS editing method based on As-Rigid-As-Possible (ARAP) deformation~\cite{sorkine2007rigid} with diffusion prior. Our method consists of two stages: an ARAP 3DGS deformation stage and a 3DGS fine-tuning stage with diffusion prior, as shown in ~\cref{fig:pipeline}. In the first stage, we transform the initial 3D Gaussians into a representative subset. We then apply the ARAP deformation directly to this subset with the user-given drag operations. The remaining 3DGS are deformed based on the interpolation of the transformed subset. However, properties such as color and opacity remain unchanged, which may lead to artifacts in the rendered images. To address this issue, in the second stage, we use an iterative optimization strategy that fine-tunes the attributes of masked 3D Gaussians with the ability of pre-trained 2D diffusion-based super-resolution model to correct the edited content and improve visual quality.
Experiments demonstrate the effectiveness of our drag-driven 3DGS editing method across various scenes and tasks. Compared to previous methods, our method significantly improves the visual quality and geometric integrity of the results. In summary, our contributions include: 
\begin{itemize} 
\item We propose ARAP-GS, a drag-driven 3DGS editing method based on ARAP deformation. To the best of our knowledge, this is the first method to apply ARAP directly to 3D Gaussians, enabling free-form 3DGS deformation without requiring additional supervision data. 
\item We introduce an effective iterative optimization strategy that utilizes 2D diffusion prior to fine-tune masked 3DGS after ARAP deformation, enhancing multi-view consistency and visual quality. 
\item Extensive experiments demonstrate our method’s ease of use and superior visual quality compared to existing methods across a variety of 3DGS scenes.
 
\end{itemize}

%% file: sec/2_relatedwork.tex
\section{Related Work}
\label{sec:relatedwork}

\subsection{3D Gaussian Splatting Editing}
NeRF~\cite{mildenhall2021nerf} and 3DGS~\cite{kerbl20233d} are both popular representations for novel view synthesis, while NeRF implicitly represents scenes in MLPs and 3DGS uses explicit representation. Although there are numerous NeRF-editing methods~\cite{liu2021editing, wang2022clip, zhuang2023dreameditor, haque2023instruct}, they cannot be applied to 3DGS due to different representations. With the growing popularity of 3DGS in computer vision, a few works also emerged to support 3DGS editing. Text-guided 3D editing methods~\cite{chen2024gaussianeditor, ye2023gaussian, wang2024gaussianeditor, palandra2024gsedit, wang2025view, wu2024gaussctrl, zhuang2024tip, sun2024gseditpro} stand out due to their ease of use in modifying 3DGS. However, they edit 3DGS with text prompts only and face a common issue: the editing results often lead to semantic changes, limiting their ability to edit geometric and textural details.

Some other methods~\cite{ma2024reconstructing, cai2024dynasurfgs, huang2024sc} enable dynamic geometry deformations based on sufficient prior knowledge extracted from monocular videos. But they require additional video supervision, and are sensitive to the camera views provided by the dataset, limiting their ability to edit more general categories. There are also some concurrent works focusing on drag-driven 3DGS editing~\cite{shen2024draggaussian,chen2024mvdrag3d}, but they both project the drag operations onto 2D images and use advanced models~\cite{pan2023drag, shi2024dragdiffusion} to edit the images, and then optimize the 3D scene based on the edited 2D images. These methods show impressive generation ability because of the advanced 2D diffusion models. However, \cite{shen2024draggaussian} uses inconsistency among different edited images, which could result in blurry regions in the optimized 3D scenes. In addition, the 2D edited result of each view may not strictly follow user's instructions \cite{chen2024mvdrag3d}. Different from these works, our method directly applies the editing operation to 3DGS and then uses the 2D diffusion prior to fine-tune the masked region, demonstrating better 3D consistency and visual quality after editing.


\subsection{Traditional 3D Deformation}
Traditional 3D deformation methods in computer graphics have contributed various optimization techniques, such as free-form deformation~\cite{sederberg1986free}, multi-resolution deformation~\cite{kobbelt2000multiresolution}, and elastic deformation~\cite{bazen2002thin}. However, they lack the ability of precise control, leading to the loss or distortion of local details during deformation. Cage-based Deformation(CBD)~\cite{sacht2015nested, ju2023mean, joshi2007harmonic, lipman2008green} has been proposed to support high-quality deformation. However, creating and manipulating cage vertices is complex and tedious, requiring high computational cost. 

Another line of work is based on physically-inspired energy, offering more promising results and intuitive manipulation~\cite{xu2007gradient, wand2007reconstruction, sieger2014constrained}. 
ARAP~\cite{sorkine2007rigid} is one of the most popular algorithms in this area, which proposes an iterative mesh deformation framework based on the principle of minimizing local deviations from rigidity, enabling intuitive and detail-preserving deformation. Its extended versions~\cite{chao2010simple, levi2014smooth, chen2017rigidity} have improved robustness using alternative energy formulations.
ARAP has been applied to various tasks~\cite{baieri2024implicit, nagata2024creation, huang2021arapreg, le2020stiff} due to its straightforward interpretation and easy optimization while satisfying handle constraints. As far as we know, we are the first to directly apply ARAP deformation to 3D Gaussians and achieve drag-driven 3DGS editing.

\subsection{2D Image Editing}
Diffusion models~\cite{ho2022cascaded, rombach2022high, saharia2022photorealistic, nichol2021improved} have shown impressive results in image generation. Diffusion Prior embedded in models like Stable Diffusion~\cite{rombach2022high} can be applied to various downstream tasks, such as image inpainting~\cite{li2022mat, wang2023imagen, yu2023inpaint, corneanu2024latentpaint}, image super-resolution~\cite{wang2024exploiting, lin2023diffbir, yu2024scaling}, and image editing~\cite{brooks2023instructpix2pix, meng2021sdedit, hertz2022prompt, avrahami2022blended, li2024blip}. While text-driven image editing based on diffusion models is user-friendly, fully conveying editing intentions through text can be challenging, leading to results that do not meet expectations.

Some methods~\cite{mou2023dragondiffusion, liu2024drag, nie2023blessing, cui2024stabledrag} have identified this limitation and begun to focus on editing image details with free-form operations. DragGAN~\cite{pan2023drag} introduces a drag-driven image editing method using point manipulation based on Generative Adversarial Networks (GANs)~\cite{goodfellow2014generative}. Given a set of handle points and target points, the image is edited by moving the handle points toward their respective target points. DragDiffusion~\cite{shi2024dragdiffusion} extends the editing framework to diffusion models, enabling more precise spatial control in image editing with significantly better generative ability. It should be noted that directly extending these drag-driven editing methods from 2D to 3D is challenging due to the 2D view ambiguity and multi-view consistency.

%% file: sec/3_method.tex
\section{Method}
\label{sec3:method}
ARAP-GS takes a pre-trained 3DGS scene as input, along with $N$ handle points $h\in\mathbb{R}^{N\times 3}$ and corresponding drag operations. The target locations $h'\in\mathbb{R}^{N\times 3}$ are then obtained from the operations. Our goal is to reposition the 3D Gaussians around the handle points to their target locations, while preserving the rigidity of the original scene and maintaining content similarity as much as possible. This process involves ARAP 3DGS deformation and appearance fine-tuning with diffusion prior. In the following, we first review the preliminary knowledge of ARAP 3D mesh deformation and 3DGS in \cref{sec3.1:preliminary}, followed by the introduction of our proposed methods, including ARAP-based 3DGS deformation in \cref{sec3.2:arap}, and fine-tuning with diffusion prior in \cref{sec3.3:stablesr}.

\begin{figure*}[t]
  \centering
  \includegraphics[width=1\linewidth]{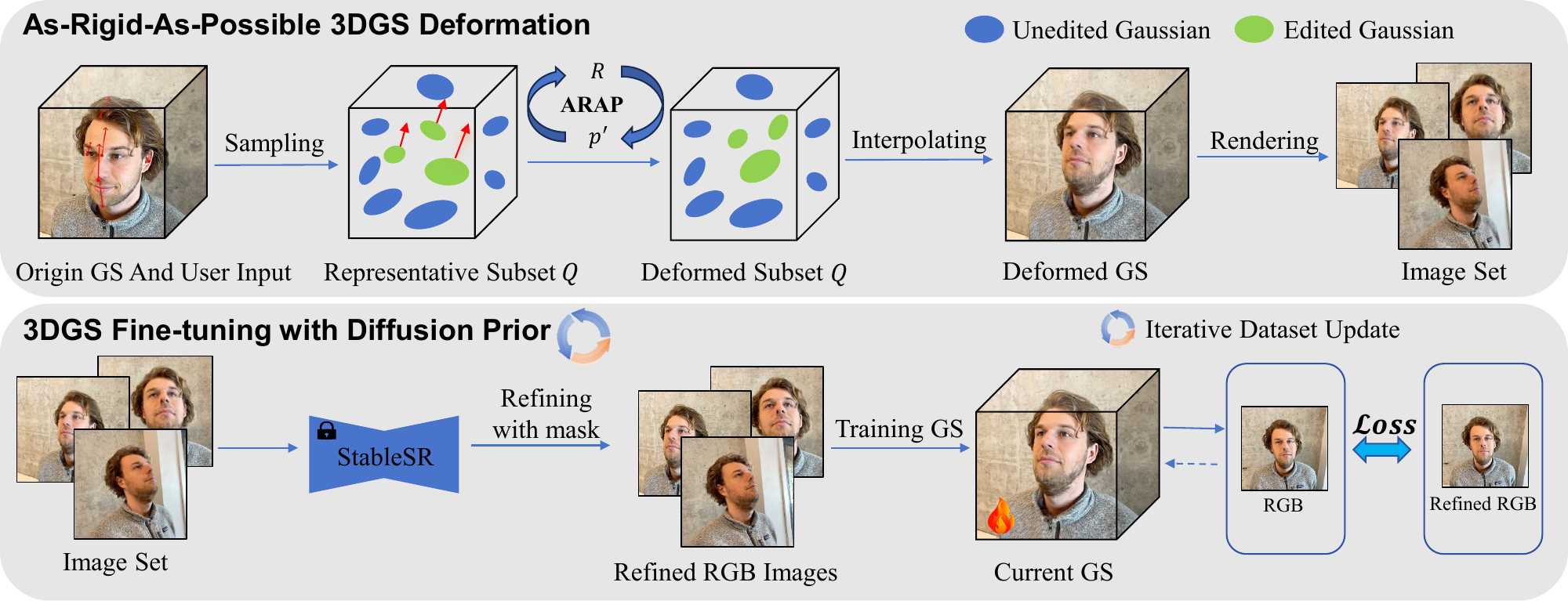}
  
   \caption{\textbf{\textit{Method overview.}} Our method is implemented in two stages. In the first stage, for geometric deformations during editing, we leverage the explicit representation of 3D Gaussians, establishing a representative subset $Q$. Then we apply the traditional ARAP deformation directly to 3D Gaussians in $Q$, and obtain the rotation matrix $R$ and new position $p'$ for each 3D Gaussian. The rotation matrices and new positions of the remaining 3D Gaussians are then inferred by interpolating those of the nearest neighbors in $Q$ (\cref{sec3.2:arap}). The deformation changes the position and covariance of the 3D Gaussians. To further update properties such as color and scale values of 3D Gaussians, we fine-tune the deformed 3D Gaussians in the second stage. Specifically, we utilize 2D diffusion prior to remove artifacts on the rendered images and iteratively optimize the 3D Gaussians based on the enhanced images (\cref{sec3.3:stablesr}).}
   \label{fig:pipeline}
\end{figure*}


\subsection{Preliminary}
\label{sec3.1:preliminary}
\textbf{ARAP 3D Mesh Deformation.} ARAP deformation~\cite{sorkine2007rigid} is a technique for manipulating 3D meshes, designed to preserve local rigidity throughout the deformation process. This objective is achieved by minimizing the following energy function:
\begin{equation}\label{eq1}
E = \sum_{i=1}^{n} w_i \sum_{j \in \mathcal{N}(i)} w_{ij} \| (p'_i - p'_j) - R_i (p_i - p_j)\|^2 ,
\end{equation}
where $\mathcal{N}(i)$ denotes the indices of the neighbors of vertex $i$, $p$ and $p'$ represent the coordinates of the vertices before and after transformation, $R$ is the rotation matrix, and $w_{i}$, $w_{ij}$ are some fixed cell and edge weights. The energy is minimized through an iterative process. Initially, $p' = p$ and $R$ is set to an identity matrix. Given a set of updated positions $ \{p'_{0}, p'_{1}, \dots, p'_{N}\}$ from the drag operation, the corresponding rotation matrices $\{R_{0}, R_{1}, \dots, R_{N}\}$ are updated by the singular value decomposition (SVD) of the covariance matrix of the edge lengths. Subsequently, the positions of other points are updated by solving the following equation:
\begin{equation}
\label{eq:p_update}
\sum_{j\in\mathcal{N}(i)}w_{ij}({p}_i'-{p}_j')=\sum_{j\in\mathcal{N}(i)}\frac{w_{ij}}{2}({R}_i+{R}_j)({p}_i-{p}_j),
\end{equation}
and the rotation matrices of other points will be updated accordingly. This process is repeated iteratively until convergence.

\noindent
\textbf{3D Gaussian Splatting.} 3D Gaussian Splatting~\cite{kerbl20233d} is an explicit representation that uses a set of anisotropic 3D Gaussians to represent the scene, denoted as $G=\{g_1, g_2, \dots, g_N\}$, where $g_i=\{\mu, \Sigma, c, \alpha\}$, $i\in\{1, \dots, N\}$. In this formulation, $\mu$ represents the center position of the Gaussian, $\Sigma$ is the 3D covariance matrix, $c$ is the color encoded with spherical harmonic coefficients, and $\alpha$ denotes opacity. To ensure that $\Sigma$ is positive semi-definite and allows independent optimization, it is decomposed as $\Sigma = RSS^{T}R^{T}$, where ${R}$ is a rotation matrix that can be easily converted to a quaternion $q$, and ${S}$ is a diagonal scaling matrix.

\subsection{As-Rigid-As-Possible 3DGS Deformation}
\label{sec3.2:arap}
To achieve geometric deformation for a scene represented in 3DGS, there are two primary challenges: (1) offering user-friendly operations, such as text prompts or sparse control points, to drive 3DGS deformations, and (2) effectively preserving object rigidity during geometric deformation of 3DGS while supporting flexible editing operations, such as translation, stretching, and rotation. In this paper, we focus on drag-driven deformation, which is user-friendly to modify complex geometric structures, allowing users to easily adjust and reshape content with minimal technical expertise. To maintain the rigidity of 3DGS during deformation, we adopt the as rigid as possible principle from the traditional ARAP 3D mesh deformation~\cite{sorkine2007rigid} and apply it directly to 3DGS for deformation. The detailed methodology is outlined below.

As shown in \cref{eq1}, the energy minimization process aims to maintain the edge length between vertices as constant as possible. Building on this insight, we believe the change of distance between 3D Gaussians should also be minimized to preserve local details after deformation. Consequently, it is natural to treat the center of 3D Gaussian as the vertex in ARAP deformation for energy minimization, extending the concept of ARAP from 3D meshes to 3DGS.

On the other hand, due to the large number of 3D Gaussians in the scene, directly solving for the transformation of all 3D Gaussians is computationally intensive and limited by both time and memory constraints. To mitigate this, we perform random sampling on the original set of 3D Gaussians to create a representative subset $Q$. We then apply K-Nearest Neighbors (KNN) to $Q$ to establish adjacency relationships and retain them for 3D Gaussians in $Q$. We initialize $p'$ and $R$ as illustrated in \cref{sec3.1:preliminary}. Subsequently, we iteratively update the rotation matrix $R_i$ and the target center $p_i'$ as described in \cref{sec3.1:preliminary} for each 3D Gaussian. When solving for $R_i$, let $\mathcal{N}(i)$ denote the indices of $p_i$’s neighbors, we have the covariance matrix $S_i=P_i D_i P_i'^T$, where $D_i \in \mathbb{R}^{|\mathcal{N}(i)|\times |\mathcal{N}(i)|}$ is a diagonal matrix composed of $w_{ij}$, here $j\in\mathcal{N}(i)$. $P_i \in \mathbb{R}^{3\times|\mathcal{N}(i)|}$ is a matrix containing the 3D Gaussian center distance $p_i-p_j$ as its columns and $P'_i$ is defined similarly. Then we perform SVD on $S_i$: $S_i=U_i\Sigma_iV_i^T$, and obtain $R_i=V_iU_i^T$. $R_i$ is used to update the centers and rotation quaternions of other 3D Gaussians, which is achieved using the same equation as~\cref{eq:p_update}. A comprehensive derivation of the mathematics is provided in ~\cite{sorkine2007rigid}.

After transforming the 3D Gaussians in $Q$, we focus on transforming the remaining 3D Gaussians through interpolation. For each Gaussian, we select the $k$ nearest 3D Gaussians from $Q$ and apply linear interpolation to obtain the corresponding rotation $q_l'$ and transformation $p_l'$:\par

\begin{equation}
p_l'=\sum_{i=1}^k w_{il} p_i'+p_l,
\label{p_interp}
\end{equation}
\begin{equation}
q_l'=\sum_{i=1}^k w_{il} q_i'\otimes q_l,
\label{q_interp}
\end{equation}
\begin{equation}
w_{il}=\frac{\exp(-\|(p_i'-p_l)\|_2)}{\sum_{i=1}^k \exp(-\|(p_i'-p_l)\|_2)},
\label{eq:weight}
\end{equation}
where $p_l$ and $q_l$ represent the original center and quaternion of the remaining 3D Gaussians indexed by $l$, $p_i'$ and $q_i'$ are those deformed in $Q$, and $\otimes$ denotes quaternion multiplication.

\subsection{3DGS Fine-tuning with Diffusion Prior}
\label{sec3.3:stablesr}
Thanks to the ARAP deformation on 3DGS, the geometric shape of the scene is well-preserved, but properties such as opacity and color remain unoptimized, which may result in slight artifacts in the rendered images. To address this issue, we leverage recent advances in diffusion-based 2D image enhancement techniques, applying the off-the-shelf StableSR~\cite{wang2024exploiting} model to refine the rendered image set. Leveraging the geometric integrity established in the first stage, StableSR can effectively remove the light artifacts and produce high-resolution images. However, we also observe that it could bring inconsistent regions across different views. Therefore, we have designed several strategies to reduce the effects of inconsistent views. 

\paragraph{Iterative Dataset Update.} To reduce the inconsistency effects of enhanced views from StableSR, we adopt an iterative dataset update (IDU) strategy from Instruct-NeRF2NeRF (I-N2N)~\cite{haque2023instruct}. Specifically, at the beginning of the fine-tuning process, we store the original captured images for each view individually. Every $t=10$ iterations, we select a subset with $n$ views and use StableSR to enhance them. These enhanced images replace the original images as the supervision for those views. This simple strategy effectively improves the 3DGS appearance while keeping the geometry properties of the scene compared to fine-tuning 3DGS with all views updated at once.\par
\paragraph{Mask-Guided 3DGS Fine-tuning.} During the 3DGS fine-tuning process, we implement additional mechanisms to further avoid inconsistent regions produced by StableSR across multiple views. The idea is to reduce the extent of corrections by masking the edited regions, thereby preserving the original view consistency as much as possible. This mask is generated by computing the displacement of GS positions, identifying points that exceed a specified threshold, and projecting them onto the camera plane. Consequently, the updated view image is represented as follows:
\begin{equation}
I_{merge} = M\odot I_{sr}+(1-M)\odot I_{gt},
\end{equation}
where $M$ denotes the mask, $I_{sr}$ denotes the enhanced image, $I_{gt}$ denotes the ground truth, and $\odot$ denotes the element-wise product. Then $I_{merge}$ is used for the computation of the loss in vanilla 3DGS.




%% file: sec/4_experiments.tex
\begin{figure*}[t]
  \centering
  \includegraphics[width=0.9\linewidth]{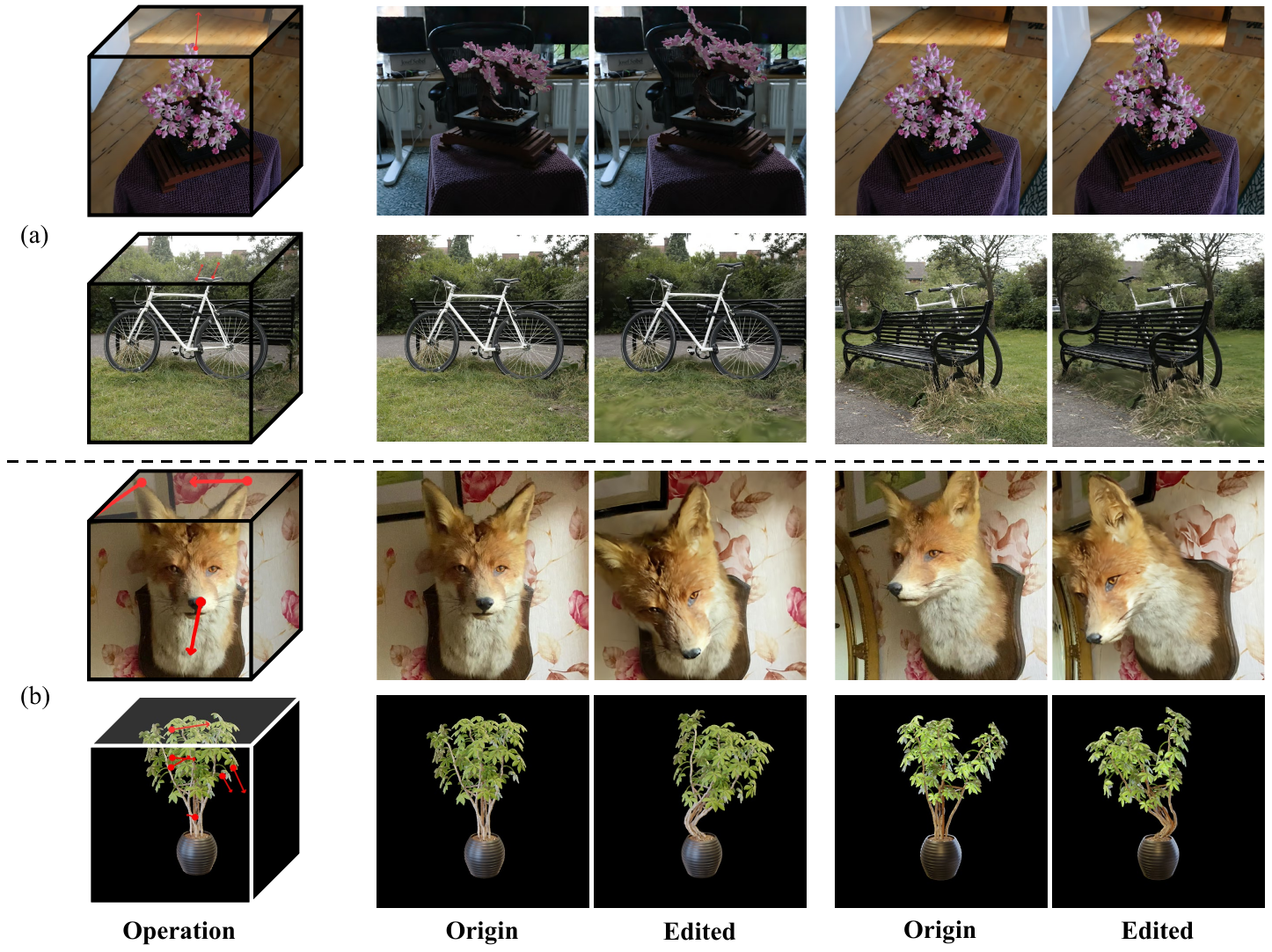}
  
   \caption{\textbf{\textit{Visual results of our method.}} The first two rows show the results of stretching, and last two rows show the results of rotation. The first column shows the dragging operation, with the red points indicating the handle points and the arrows indicating the dragging directions. "Original" and "Edited" denote the rendering results before and after editing.}
   
   \label{fig:result}
\end{figure*}
\section{Experiments}
\label{sec4:exp}


\subsection{Experimental Setup}

\textbf{Dataset.} We use 10 scenes for evaluation, varying from real scenes to a single object. These scenes are collected from ~\cite{haque2023instruct,barron2022mip,mildenhall2021nerf,muller2022instant}.

\noindent
\textbf{Implementation Details.} We conduct our experiments on a single RTX 3090. The size of the original scene ranges from 500k to 3000k Gaussians. Considering that some scenes have Gaussians far from those to be edited, we manually set a mask to filter them. The size of the representative Gaussian subset is $N=16384$, with KNN for constructing neighbor relationships set to $k=32$, and the number of neighbors used in the interpolation is 8. The maximum number of ARAP iterations is 16. During the iterative dataset update, we follow the I-N2N~\cite{haque2023instruct} setup, updating one viewpoint every 10 iterations. The number of 3DGS optimization iterations ranges from 800 to 2000 depending on the number of initial Gaussians. More details about the sensitive parameters can be found in the supplementary material.\par

\noindent
\textbf{Baselines.} To the best of our knowledge, there are no publicly available methods for drag-driven 3DGS editing. Therefore, we choose two representative text-driven 3DGS editing methods, I-N2N~\cite{haque2023instruct} and GaussianEditor (GSEditor)~\cite{chen2024gaussianeditor} for comparison. Additionally, we select two representative drag-driven 2D image editing methods, DragDiffusion (DragDiff) ~\cite{shi2024dragdiffusion} and SDEDrag~\cite{nie2023blessing}, as baselines. Specifically, we apply the 2D editing methods on the rendered views from 3DGS, and then use the edited results to fine-tune parameters of 3DGS.

\noindent
\textbf{Running Time.} Our editing time depends on the number of optimization iterations and the scene size. For a scene with around 500k Gaussians to be edited, ARAP takes approximately 6 minutes, while 1000 iterations of fine-tuning optimization take about 5 minutes. In comparison, I-N2N needs about an hour and GSEditor needs around 20 minutes. For each viewpoint, DragDiff takes an average of 5 minutes, while SDEDrag takes an average of 15 minutes. Their total time depends on the number of viewpoints, with a minimum of 2 hours required for the fastest case. 

\noindent
\textbf{Evaluation Metrics.} We employ Dragging Accuracy Index (DAI)~\cite{zhang2024gooddrag} for quantitative evaluation. DAI is defined as follows: 
\begin{equation}
DAI = \frac{1}{n}
\sum_{i=1}^n
\sum_{j=1}^{m}
\frac{\left\|\Omega({I}_{o,i},{p}_{i,j},\gamma)-\Omega({I}_{e,i},{q}_{i,j},\gamma)\right\|_2^2}{(1+2\gamma)^2},
\end{equation}
where $n$ denotes the number of views, $m$ denotes the number of handle points, $I_{o,i}$ and $I_{e,i}$ denote the image before and after editing on $i$-th view, $p_{i,j}$ and $q_{i,j}$ denote the handle and target points projected onto the camera plane, and $\Omega({I}_{o,i},{p}_{i,j},\gamma)$ denotes a square patch of ${I}_{o,i}$ centered at ${p}_{i,j}$ with size $\gamma$. We randomly select $n=10$ views for evaluation, and set $\gamma=\{1,5,10,20\}$ to average different levels of context. DAI measures the effectiveness of a method in transferring source content to a target point. Considering that the quality of editing is closely related to user's perception, we further conduct both user study and GPT evaluation on the editing
results. We collect votes from 60 participants in the user study, and the edited results from different methods are randomly placed in each question. Since \cite{wu2024gpt} proposed that GPT's ability to evaluate images aligns with human perception, we present images to GPT-4o and ask it to assess the effectiveness of the edits, receiving a score on a scale of 1 to 10. The average score of each method is also reported in ~\cref{tab:all}.

\subsection{Qualitative Results} 
\begin{figure*}[t]
  \centering
  \includegraphics[width=1\linewidth]{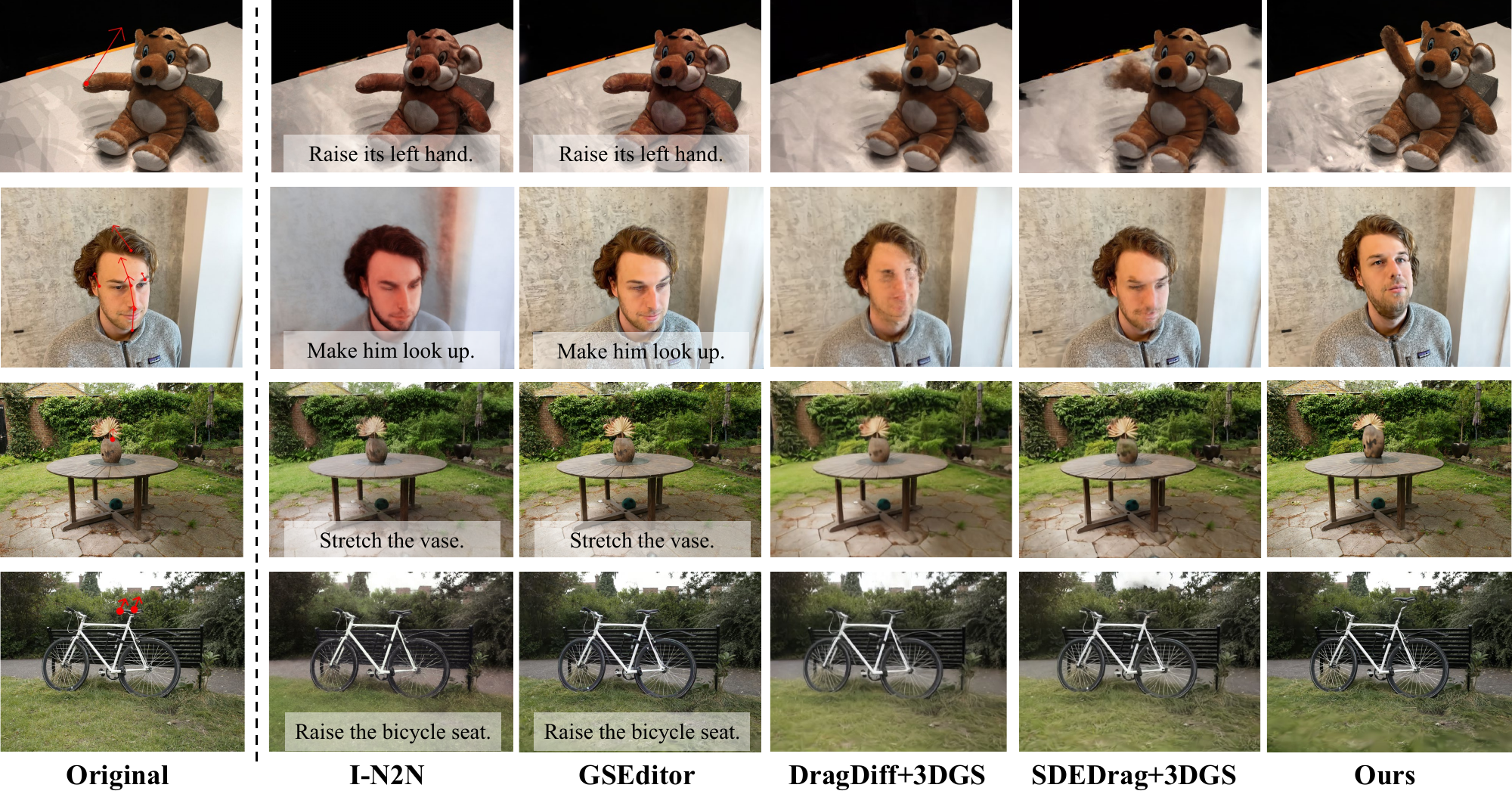}
  
   \caption{\textbf{\textit{Qualitative comparison results.}}  We compare our method with state-of-the-art methods including I-N2N~\cite{haque2023instruct}, GaussianEditor~\cite{chen2024gaussianeditor}, DragDiffusion~\cite{shi2024dragdiffusion} and SDEDrag~\cite{nie2023blessing}. Compared to these methods, our method achieves more accurate geometric deformation and better multi-view consistency, delivering higher visual quality results.}
   \label{fig:eval}
\end{figure*}

We present the edited results of our method from different drag operations in \cref{fig:result}. As shown in the first two rows, where red arrows indicate the stretching direction. It is obvious that our method achieves stretching in line with the user’s intent, maintaining visual consistency across different viewpoints. The last two rows in \cref{fig:result} present the rotation transformation, and red arrows indicate the rotation direction. By annotating a few handle points and specifying the rotation angle, the 3D scene can achieve visual plausible rotation deformation. 


As shown in \cref{fig:eval}, although I-N2N and GSEditor can perform text-based 3D editing, they have difficulty handling complex action commands, such as "look up" or "stretch", due to the limited text understanding ability. As a result, they often fail to perform accurate geometric transformations. This limitation highlights the superiority of our geometric 3DGS deformation strategy. 
While DragDiff and SDEDrag achieve satisfactory geometric edits in 2D, they fail to produce consistent edits when applied to 3D due to a lack of spatial coherence across viewpoints. This leads to artifacts in 3DGS after optimization. In contrast, our method maintains better appearance after deformation due to the direct ARAP 3DGS deformation and view-consistent image enhancement with diffusion prior. More qualitative results are presented in the supplementary material.

\subsection{Quantitative Results}
The quantitative comparison results are shown in ~\cref{tab:all}. Our method achieves the highest scores across all metrics compared to other methods in the drag-driven 3DGS editing task. The DAI scores indicate that our method effectively maintains content similarity between the source and target positions after deformation. The user study demonstrates that participants significantly prefer our method over others, which fall short in the drag-driven 3DGS editing task. Additionally, the GTP-4o scores further validate that our results achieve better results and are more consistency with input deformation.

\subsection{Ablation Study}
We mainly discuss the effects of the two main designs in our method: ARAP deformation and diffusion prior. More ablation studies about other designs are included in the supplementary material.
\begin{table*}[!htb]
    \centering
    \begin{tabular*}{\textwidth}{@{\extracolsep{\fill}}c|ccccc}
    \toprule
    \ &I-N2N\ &GSEditor\ &DragDiff+3DGS\ &SDEDrag+3DGS\ &Ours\ \\
    \midrule
    DAI($\downarrow$)\ &0.3249\ &0.2994\ &0.2811\ &0.3037\ &\textbf{0.0968}\ \\
    User Votes($\uparrow$)\ &5.1\%	\ &7.8\% \ &5.1\% \ &4.2\% \ &\textbf{77.8\%}\ \\
    GTP-4o($\uparrow$)\ &4.600\ &5.467\ &4.733\ &5.867\ &\textbf{8.067}\ \\
    \bottomrule
    \end{tabular*}
    \caption{\textbf{\textit{Quantitative evaluation.}} We compare our method with state-of-the-art methods including Instruct-NeRF2NeRF~\cite{haque2023instruct}, GaussianEditor~\cite{chen2024gaussianeditor}, DragDiffusion~\cite{shi2024dragdiffusion} and SDEDrag~\cite{nie2023blessing}. Compared to these methods, our method outperforms in all evaluation metrics.}
    \label{tab:all}
\end{table*}

\begin{figure}[!h]
  \centering
  \includegraphics[width=1\linewidth]{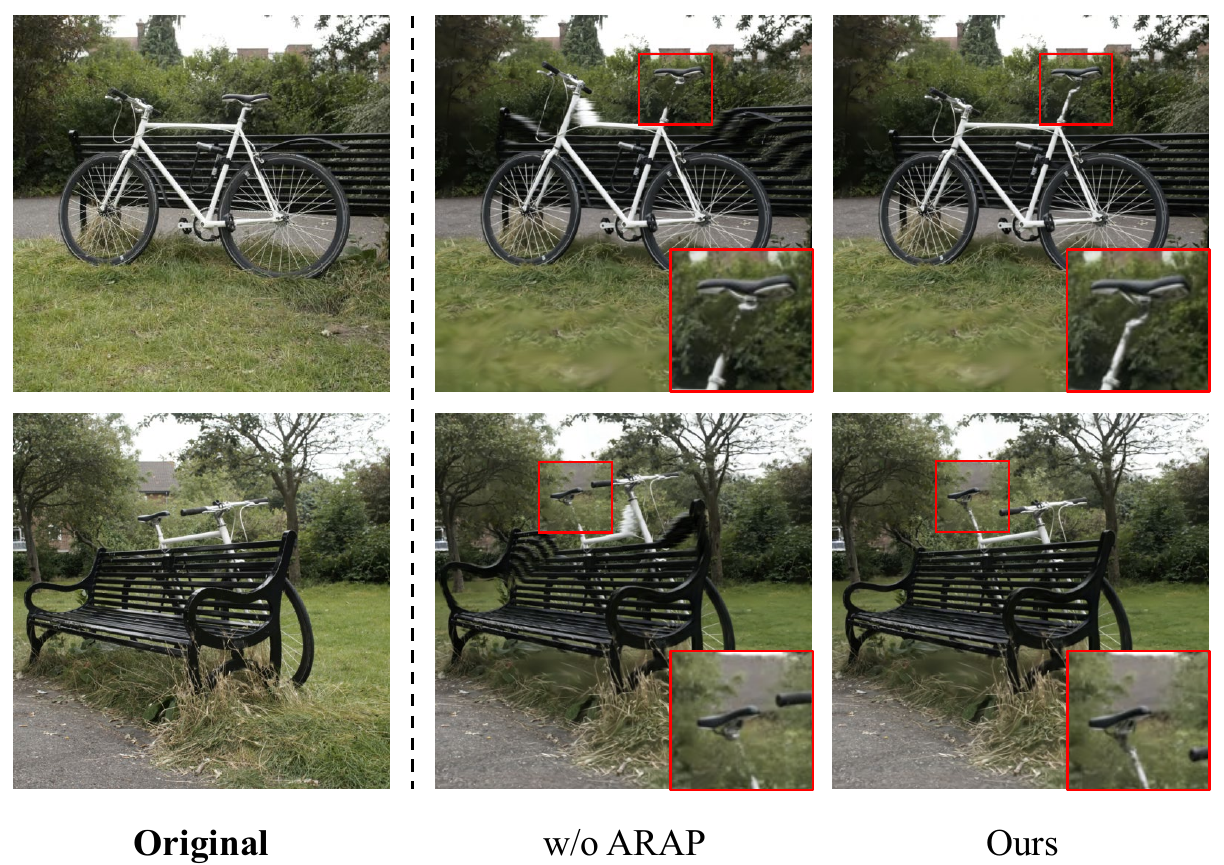}
  
   \caption{\textbf{\textit{Effectiveness of ARAP Deformation.}} We select two views to compare the results with and without ARAP Deformation. We magnify the deformed region to highlight our method's ability to preserve the geometric structure of the scene after deformation.}
   \label{fig:abla1}
\end{figure}

\noindent
\textbf{Effectiveness of ARAP Deformation.} To verify the effectiveness of the ARAP deformation, we compare it with direct interpolation, where the other Gaussians are linearly interpolated directly based on the nearest input handle points and deformation, using the weight defined in ~\cref{eq:weight} . As shown in \cref{fig:abla1}, our method achieves more precise modifications and better preservation of the geometric information of the original model. In contrast, using direct interpolation not only fails to deform the Gaussians correctly, resulting in broken parts, but also causes unexpected changes in other regions. Additionally, a direct dragging and interpolation strategy struggles to perform rotation on Gaussians.

\begin{figure}[!h]
  \centering
  \includegraphics[width=1\linewidth]{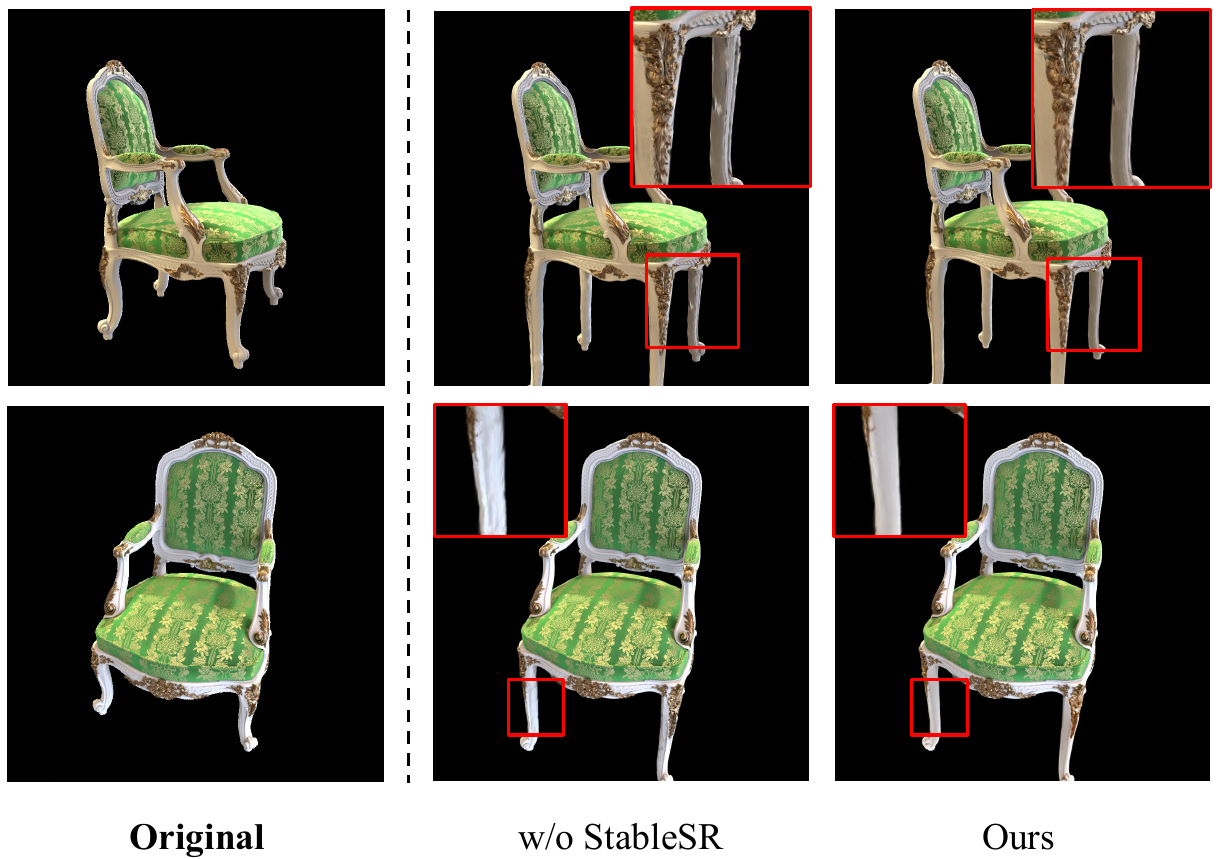}
  
   \caption{\textbf{\textit{Effectiveness of Diffusion Prior.}} We select two views to compare the results with and without Diffusion Prior. We magnify the deformed region to highlight our method's ability to maintain high-fidelity rendering quality after deformation.}
   \label{fig:abla2}
\end{figure}

\noindent
\textbf{Effectiveness of Diffusion Prior.} We show the results without diffusion prior in \cref{fig:abla2}. It is obvious that, in addition to geometric 3DGS deformation, incorporating a super-resolution model further enhances the appearance of Gaussians by fine-tuning the 3DGS parameters with enhanced images, resulting in improved quality with reduced artifacts.

%% file: sec/5_conclusion.tex
\section{Conclusion, Limitation and Future Work}
We propose ARAP-GS, a drag-driven ARAP 3DGS editing method with diffusion prior. We directly apply ARAP deformation to 3D Gaussians, enabling free-form geometric deformation of 3DGS without additional supervision data. Furthermore, we leverage the content-enhancing capabilities of diffusion prior to refine the rendered images and fine-tune the appearance of 3DGS. To maintain view consistency and geometric integrity during the fine-tuning process, we implement several strategies, such as iterative dataset updates and mask-guided 3DGS fine-tuning. Extensive experiments demonstrate that our method achieves high-quality drag-driven 3DGS editing and shows clear superiority over other methods across various scenes.\par
\begin{figure}[!h]
  \centering
  \includegraphics[width=1\linewidth]{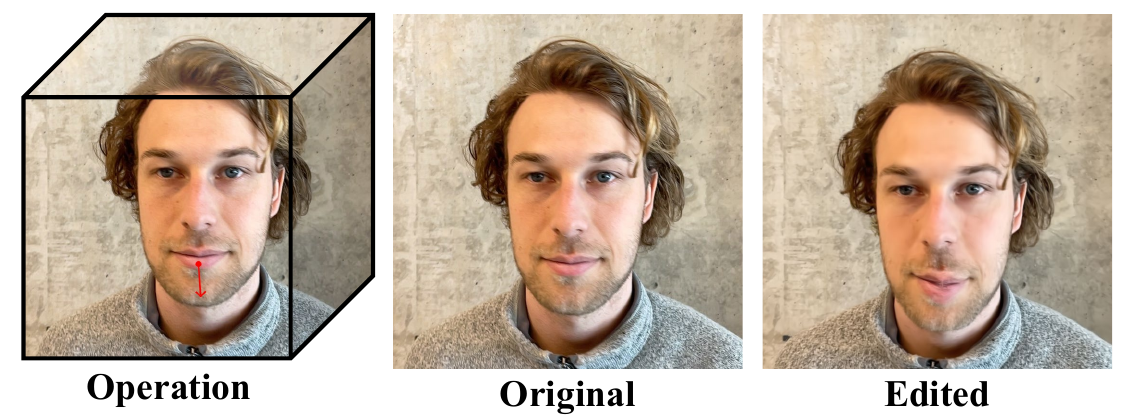}
  
   \caption{\textbf{\textit{Failure case of our method.}} When attempting to make the man open his mouth, the Gaussians in the mouth region maintain their original structure and do not tear to create holes, leading to less satisfactory results in this case.}
   \label{fig:failcase}
\end{figure}

\noindent
\textbf{Limitation.} Our method preserves the topological structure of the scene with minimal disruption, so some desired editing results, as illustrated in ~\cref{fig:failcase}, are not supported by our method.

\noindent
\textbf{Future work.} Although our method has made pioneering contributions to the drag-driven 3DGS editing task, we have not yet considered directly modifying other properties of 3D Gaussians, such as scaling and color, during ARAP deformation. In the future, we plan to explore additional 3DGS property transformations during ARAP deformation. Furthermore, while we improve image quality using diffusion prior, we have not fully leveraged the generative capabilities of diffusion models. Future work could incorporate other diffusion models, like ControlNet~\cite{zhang2023adding}, to enhance the generative capabilities of the proposed method. It would also be interesting to explore other traditional mesh operations for 3DGS deformation, such as handle-based~\cite{liu2021deepmetahandles} and skeleton-based~\cite{capell2002interactive} deformation.

%% file: main.bbl
\begin{thebibliography}{72}
\providecommand{\natexlab}[1]{#1}
\providecommand{\url}[1]{\texttt{#1}}
\expandafter\ifx\csname urlstyle\endcsname\relax
  \providecommand{\doi}[1]{doi: #1}\else
  \providecommand{\doi}{doi: \begingroup \urlstyle{rm}\Url}\fi

\bibitem[Avrahami et~al.(2022)Avrahami, Lischinski, and Fried]{avrahami2022blended}
Omri Avrahami, Dani Lischinski, and Ohad Fried.
\newblock Blended diffusion for text-driven editing of natural images.
\newblock In \emph{Proceedings of the IEEE/CVF conference on computer vision and pattern recognition}, pages 18208--18218, 2022.

\bibitem[Bai et~al.(2024)Bai, Huang, Guo, Gong, Li, and Guo]{bai2024360}
Jiayang Bai, Letian Huang, Jie Guo, Wen Gong, Yuanqi Li, and Yanwen Guo.
\newblock 360-gs: Layout-guided panoramic gaussian splatting for indoor roaming.
\newblock \emph{arXiv preprint arXiv:2402.00763}, 2024.

\bibitem[Baieri et~al.(2024)Baieri, Maggioli, L{\"a}hner, Melzi, and Rodol{\`a}]{baieri2024implicit}
Daniele Baieri, Filippo Maggioli, Zorah L{\"a}hner, Simone Melzi, and Emanuele Rodol{\`a}.
\newblock Implicit-arap: Efficient handle-guided deformation of high-resolution meshes and neural fields via local patch meshing.
\newblock \emph{arXiv preprint arXiv:2405.12895}, 2024.

\bibitem[Barron et~al.(2022)Barron, Mildenhall, Verbin, Srinivasan, and Hedman]{barron2022mip}
Jonathan~T Barron, Ben Mildenhall, Dor Verbin, Pratul~P Srinivasan, and Peter Hedman.
\newblock Mip-nerf 360: Unbounded anti-aliased neural radiance fields.
\newblock In \emph{Proceedings of the IEEE/CVF conference on computer vision and pattern recognition}, pages 5470--5479, 2022.

\bibitem[Bazen and Gerez(2002)]{bazen2002thin}
Asker~M Bazen and Sabih~H Gerez.
\newblock Thin-plate spline modelling of elastic deformations in fingerprints.
\newblock In \emph{Proceedings of 3rd IEEE Benelux Signal Processing Symposium}. Citeseer, 2002.

\bibitem[Brooks et~al.(2023)Brooks, Holynski, and Efros]{brooks2023instructpix2pix}
Tim Brooks, Aleksander Holynski, and Alexei~A Efros.
\newblock Instructpix2pix: Learning to follow image editing instructions.
\newblock In \emph{Proceedings of the IEEE/CVF Conference on Computer Vision and Pattern Recognition}, pages 18392--18402, 2023.

\bibitem[Cai et~al.(2024)Cai, Ye, Ye, He, and Chen]{cai2024dynasurfgs}
Weiwei Cai, Weicai Ye, Peng Ye, Tong He, and Tao Chen.
\newblock Dynasurfgs: Dynamic surface reconstruction with planar-based gaussian splatting.
\newblock \emph{arXiv preprint arXiv:2408.13972}, 2024.

\bibitem[Capell et~al.(2002)Capell, Green, Curless, Duchamp, and Popovi{\'c}]{capell2002interactive}
Steve Capell, Seth Green, Brian Curless, Tom Duchamp, and Zoran Popovi{\'c}.
\newblock Interactive skeleton-driven dynamic deformations.
\newblock \emph{ACM transactions on graphics (TOG)}, 21\penalty0 (3):\penalty0 586--593, 2002.

\bibitem[Chao et~al.(2010)Chao, Pinkall, Sanan, and Schr{\"o}der]{chao2010simple}
Isaac Chao, Ulrich Pinkall, Patrick Sanan, and Peter Schr{\"o}der.
\newblock A simple geometric model for elastic deformations.
\newblock \emph{ACM transactions on graphics (TOG)}, 29\penalty0 (4):\penalty0 1--6, 2010.

\bibitem[Chen et~al.(2024{\natexlab{a}})Chen, Lan, Chen, Zhou, and Pan]{chen2024mvdrag3d}
Honghua Chen, Yushi Lan, Yongwei Chen, Yifan Zhou, and Xingang Pan.
\newblock Mvdrag3d: Drag-based creative 3d editing via multi-view generation-reconstruction priors.
\newblock \emph{arXiv preprint arXiv:2410.16272}, 2024{\natexlab{a}}.

\bibitem[Chen et~al.(2017)Chen, Gao, Lai, and Xia]{chen2017rigidity}
Shu-Yu Chen, Lin Gao, Yu-Kun Lai, and Shihong Xia.
\newblock Rigidity controllable as-rigid-as-possible shape deformation.
\newblock \emph{Graphical Models}, 91:\penalty0 13--21, 2017.

\bibitem[Chen et~al.(2024{\natexlab{b}})Chen, Chen, Zhang, Wang, Yang, Wang, Cai, Yang, Liu, and Lin]{chen2024gaussianeditor}
Yiwen Chen, Zilong Chen, Chi Zhang, Feng Wang, Xiaofeng Yang, Yikai Wang, Zhongang Cai, Lei Yang, Huaping Liu, and Guosheng Lin.
\newblock Gaussianeditor: Swift and controllable 3d editing with gaussian splatting.
\newblock In \emph{Proceedings of the IEEE/CVF Conference on Computer Vision and Pattern Recognition}, pages 21476--21485, 2024{\natexlab{b}}.

\bibitem[Corneanu et~al.(2024)Corneanu, Gadde, and Martinez]{corneanu2024latentpaint}
Ciprian Corneanu, Raghudeep Gadde, and Aleix~M Martinez.
\newblock Latentpaint: Image inpainting in latent space with diffusion models.
\newblock In \emph{Proceedings of the IEEE/CVF Winter Conference on Applications of Computer Vision}, pages 4334--4343, 2024.

\bibitem[Cui et~al.(2024)Cui, Zhao, Zhang, Cao, Ma, and Wang]{cui2024stabledrag}
Yutao Cui, Xiaotong Zhao, Guozhen Zhang, Shengming Cao, Kai Ma, and Limin Wang.
\newblock Stabledrag: Stable dragging for point-based image editing.
\newblock \emph{arXiv preprint arXiv:2403.04437}, 2024.

\bibitem[Goodfellow et~al.(2014)Goodfellow, Pouget-Abadie, Mirza, Xu, Warde-Farley, Ozair, Courville, and Bengio]{goodfellow2014generative}
Ian Goodfellow, Jean Pouget-Abadie, Mehdi Mirza, Bing Xu, David Warde-Farley, Sherjil Ozair, Aaron Courville, and Yoshua Bengio.
\newblock Generative adversarial nets.
\newblock \emph{Advances in neural information processing systems}, 27, 2014.

\bibitem[Haque et~al.(2023)Haque, Tancik, Efros, Holynski, and Kanazawa]{haque2023instruct}
Ayaan Haque, Matthew Tancik, Alexei~A Efros, Aleksander Holynski, and Angjoo Kanazawa.
\newblock Instruct-nerf2nerf: Editing 3d scenes with instructions.
\newblock In \emph{Proceedings of the IEEE/CVF International Conference on Computer Vision}, pages 19740--19750, 2023.

\bibitem[Hertz et~al.(2022)Hertz, Mokady, Tenenbaum, Aberman, Pritch, and Cohen-Or]{hertz2022prompt}
Amir Hertz, Ron Mokady, Jay Tenenbaum, Kfir Aberman, Yael Pritch, and Daniel Cohen-Or.
\newblock Prompt-to-prompt image editing with cross attention control.
\newblock \emph{arXiv preprint arXiv:2208.01626}, 2022.

\bibitem[Ho et~al.(2022)Ho, Saharia, Chan, Fleet, Norouzi, and Salimans]{ho2022cascaded}
Jonathan Ho, Chitwan Saharia, William Chan, David~J Fleet, Mohammad Norouzi, and Tim Salimans.
\newblock Cascaded diffusion models for high fidelity image generation.
\newblock \emph{Journal of Machine Learning Research}, 23\penalty0 (47):\penalty0 1--33, 2022.

\bibitem[Huang and Yu(2024)]{huang2024gsdeformer}
Jiajun Huang and Hongchuan Yu.
\newblock Gsdeformer: Direct cage-based deformation for 3d gaussian splatting.
\newblock \emph{arXiv preprint arXiv:2405.15491}, 2024.

\bibitem[Huang et~al.(2021)Huang, Huang, Sun, Zhang, Jiang, and Bajaj]{huang2021arapreg}
Qixing Huang, Xiangru Huang, Bo Sun, Zaiwei Zhang, Junfeng Jiang, and Chandrajit Bajaj.
\newblock Arapreg: An as-rigid-as possible regularization loss for learning deformable shape generators.
\newblock In \emph{Proceedings of the IEEE/CVF international conference on computer vision}, pages 5815--5825, 2021.

\bibitem[Huang et~al.(2024)Huang, Sun, Yang, Lyu, Cao, and Qi]{huang2024sc}
Yi-Hua Huang, Yang-Tian Sun, Ziyi Yang, Xiaoyang Lyu, Yan-Pei Cao, and Xiaojuan Qi.
\newblock Sc-gs: Sparse-controlled gaussian splatting for editable dynamic scenes.
\newblock In \emph{Proceedings of the IEEE/CVF Conference on Computer Vision and Pattern Recognition}, pages 4220--4230, 2024.

\bibitem[Joshi et~al.(2007)Joshi, Meyer, DeRose, Green, and Sanocki]{joshi2007harmonic}
Pushkar Joshi, Mark Meyer, Tony DeRose, Brian Green, and Tom Sanocki.
\newblock Harmonic coordinates for character articulation.
\newblock \emph{ACM transactions on graphics (TOG)}, 26\penalty0 (3):\penalty0 71--es, 2007.

\bibitem[Ju et~al.(2023)Ju, Schaefer, and Warren]{ju2023mean}
Tao Ju, Scott Schaefer, and Joe Warren.
\newblock Mean value coordinates for closed triangular meshes.
\newblock In \emph{Seminal Graphics Papers: Pushing the Boundaries, Volume 2}, pages 223--228. 2023.

\bibitem[Kerbl et~al.(2023)Kerbl, Kopanas, Leimk{\"u}hler, and Drettakis]{kerbl20233d}
Bernhard Kerbl, Georgios Kopanas, Thomas Leimk{\"u}hler, and George Drettakis.
\newblock 3d gaussian splatting for real-time radiance field rendering.
\newblock \emph{ACM Transactions on Graphics}, 42\penalty0 (4):\penalty0 1--14, 2023.

\bibitem[Kobbelt et~al.(2000)Kobbelt, Bareuther, and Seidel]{kobbelt2000multiresolution}
Leif~P Kobbelt, Thilo Bareuther, and Hans-Peter Seidel.
\newblock Multiresolution shape deformations for meshes with dynamic vertex connectivity.
\newblock In \emph{Computer Graphics Forum}, pages 249--260. Wiley Online Library, 2000.

\bibitem[Le~Vaou et~al.(2020)Le~Vaou, L{\'e}on, Hahmann, Masfrand, and Mika]{le2020stiff}
Youna Le~Vaou, Jean-Claude L{\'e}on, Stefanie Hahmann, St{\'e}phane Masfrand, and Matthieu Mika.
\newblock As-stiff-as-needed surface deformationcombining arap energy with an anisotropic material.
\newblock \emph{Computer-Aided Design}, 121:\penalty0 102803, 2020.

\bibitem[Lee et~al.(2024)Lee, Won, Jung, Bae, and Jeon]{lee2024fully}
Junoh Lee, Chang-Yeon Won, Hyunjun Jung, Inhwan Bae, and Hae-Gon Jeon.
\newblock Fully explicit dynamic gaussian splatting.
\newblock \emph{arXiv preprint arXiv:2410.15629}, 2024.

\bibitem[Levi and Gotsman(2014)]{levi2014smooth}
Zohar Levi and Craig Gotsman.
\newblock Smooth rotation enhanced as-rigid-as-possible mesh animation.
\newblock \emph{IEEE transactions on visualization and computer graphics}, 21\penalty0 (2):\penalty0 264--277, 2014.

\bibitem[Li et~al.(2024)Li, Li, and Hoi]{li2024blip}
Dongxu Li, Junnan Li, and Steven Hoi.
\newblock Blip-diffusion: Pre-trained subject representation for controllable text-to-image generation and editing.
\newblock \emph{Advances in Neural Information Processing Systems}, 36, 2024.

\bibitem[Li et~al.(2022)Li, Lin, Zhou, Qi, Wang, and Jia]{li2022mat}
Wenbo Li, Zhe Lin, Kun Zhou, Lu Qi, Yi Wang, and Jiaya Jia.
\newblock Mat: Mask-aware transformer for large hole image inpainting.
\newblock In \emph{Proceedings of the IEEE/CVF conference on computer vision and pattern recognition}, pages 10758--10768, 2022.

\bibitem[Lin et~al.(2023)Lin, He, Chen, Lyu, Dai, Yu, Ouyang, Qiao, and Dong]{lin2023diffbir}
Xinqi Lin, Jingwen He, Ziyan Chen, Zhaoyang Lyu, Bo Dai, Fanghua Yu, Wanli Ouyang, Yu Qiao, and Chao Dong.
\newblock Diffbir: Towards blind image restoration with generative diffusion prior.
\newblock \emph{arXiv preprint arXiv:2308.15070}, 2023.

\bibitem[Lipman et~al.(2008)Lipman, Levin, and Cohen-Or]{lipman2008green}
Yaron Lipman, David Levin, and Daniel Cohen-Or.
\newblock Green coordinates.
\newblock \emph{ACM transactions on graphics (TOG)}, 27\penalty0 (3):\penalty0 1--10, 2008.

\bibitem[Liu et~al.(2024)Liu, Xu, Yang, Zeng, and He]{liu2024drag}
Haofeng Liu, Chenshu Xu, Yifei Yang, Lihua Zeng, and Shengfeng He.
\newblock Drag your noise: Interactive point-based editing via diffusion semantic propagation.
\newblock In \emph{Proceedings of the IEEE/CVF Conference on Computer Vision and Pattern Recognition}, pages 6743--6752, 2024.

\bibitem[Liu et~al.(2021{\natexlab{a}})Liu, Sung, Mech, and Su]{liu2021deepmetahandles}
Minghua Liu, Minhyuk Sung, Radomir Mech, and Hao Su.
\newblock Deepmetahandles: Learning deformation meta-handles of 3d meshes with biharmonic coordinates.
\newblock In \emph{Proceedings of the IEEE/CVF Conference on Computer Vision and Pattern Recognition}, pages 12--21, 2021{\natexlab{a}}.

\bibitem[Liu et~al.(2021{\natexlab{b}})Liu, Zhang, Zhang, Zhang, Zhu, and Russell]{liu2021editing}
Steven Liu, Xiuming Zhang, Zhoutong Zhang, Richard Zhang, Jun-Yan Zhu, and Bryan Russell.
\newblock Editing conditional radiance fields.
\newblock In \emph{Proceedings of the IEEE/CVF international conference on computer vision}, pages 5773--5783, 2021{\natexlab{b}}.

\bibitem[Lu et~al.(2024)Lu, Guo, Hui, Chen, Yang, Tang, Zhu, and Dai]{lu20243d}
Zhicheng Lu, Xiang Guo, Le Hui, Tianrui Chen, Min Yang, Xiao Tang, Feng Zhu, and Yuchao Dai.
\newblock 3d geometry-aware deformable gaussian splatting for dynamic view synthesis.
\newblock In \emph{Proceedings of the IEEE/CVF Conference on Computer Vision and Pattern Recognition}, pages 8900--8910, 2024.

\bibitem[Ma et~al.(2024)Ma, Luo, and Yang]{ma2024reconstructing}
Shaojie Ma, Yawei Luo, and Yi Yang.
\newblock Reconstructing and simulating dynamic 3d objects with mesh-adsorbed gaussian splatting.
\newblock \emph{arXiv preprint arXiv:2406.01593}, 2024.

\bibitem[Meng et~al.(2021)Meng, He, Song, Song, Wu, Zhu, and Ermon]{meng2021sdedit}
Chenlin Meng, Yutong He, Yang Song, Jiaming Song, Jiajun Wu, Jun-Yan Zhu, and Stefano Ermon.
\newblock Sdedit: Guided image synthesis and editing with stochastic differential equations.
\newblock \emph{arXiv preprint arXiv:2108.01073}, 2021.

\bibitem[Mildenhall et~al.(2021)Mildenhall, Srinivasan, Tancik, Barron, Ramamoorthi, and Ng]{mildenhall2021nerf}
Ben Mildenhall, Pratul~P Srinivasan, Matthew Tancik, Jonathan~T Barron, Ravi Ramamoorthi, and Ren Ng.
\newblock Nerf: Representing scenes as neural radiance fields for view synthesis.
\newblock \emph{Communications of the ACM}, 65\penalty0 (1):\penalty0 99--106, 2021.

\bibitem[Mou et~al.(2023)Mou, Wang, Song, Shan, and Zhang]{mou2023dragondiffusion}
Chong Mou, Xintao Wang, Jiechong Song, Ying Shan, and Jian Zhang.
\newblock Dragondiffusion: Enabling drag-style manipulation on diffusion models.
\newblock \emph{arXiv preprint arXiv:2307.02421}, 2023.

\bibitem[M{\"u}ller et~al.(2022)M{\"u}ller, Evans, Schied, and Keller]{muller2022instant}
Thomas M{\"u}ller, Alex Evans, Christoph Schied, and Alexander Keller.
\newblock Instant neural graphics primitives with a multiresolution hash encoding.
\newblock \emph{ACM transactions on graphics (TOG)}, 41\penalty0 (4):\penalty0 1--15, 2022.

\bibitem[Nagata and Imahori(2024)]{nagata2024creation}
Yuichi Nagata and Shinji Imahori.
\newblock Creation of dihedral escher-like tilings based on as-rigid-as-possible deformation.
\newblock \emph{ACM Transactions on Graphics}, 43\penalty0 (2):\penalty0 1--18, 2024.

\bibitem[Nichol and Dhariwal(2021)]{nichol2021improved}
Alexander~Quinn Nichol and Prafulla Dhariwal.
\newblock Improved denoising diffusion probabilistic models.
\newblock In \emph{International conference on machine learning}, pages 8162--8171. PMLR, 2021.

\bibitem[Nie et~al.(2023)Nie, Guo, Lu, Zhou, Zheng, and Li]{nie2023blessing}
Shen Nie, Hanzhong~Allan Guo, Cheng Lu, Yuhao Zhou, Chenyu Zheng, and Chongxuan Li.
\newblock The blessing of randomness: Sde beats ode in general diffusion-based image editing.
\newblock \emph{arXiv preprint arXiv:2311.01410}, 2023.

\bibitem[Palandra et~al.(2024)Palandra, Sanchietti, Baieri, and Rodol{\`a}]{palandra2024gsedit}
Francesco Palandra, Andrea Sanchietti, Daniele Baieri, and Emanuele Rodol{\`a}.
\newblock Gsedit: Efficient text-guided editing of 3d objects via gaussian splatting.
\newblock \emph{arXiv preprint arXiv:2403.05154}, 2024.

\bibitem[Pan et~al.(2023)Pan, Tewari, Leimk{\"u}hler, Liu, Meka, and Theobalt]{pan2023drag}
Xingang Pan, Ayush Tewari, Thomas Leimk{\"u}hler, Lingjie Liu, Abhimitra Meka, and Christian Theobalt.
\newblock Drag your gan: Interactive point-based manipulation on the generative image manifold.
\newblock In \emph{ACM SIGGRAPH 2023 Conference Proceedings}, pages 1--11, 2023.

\bibitem[Rombach et~al.(2022)Rombach, Blattmann, Lorenz, Esser, and Ommer]{rombach2022high}
Robin Rombach, Andreas Blattmann, Dominik Lorenz, Patrick Esser, and Bj{\"o}rn Ommer.
\newblock High-resolution image synthesis with latent diffusion models.
\newblock In \emph{Proceedings of the IEEE/CVF conference on computer vision and pattern recognition}, pages 10684--10695, 2022.

\bibitem[Sacht et~al.(2015)Sacht, Vouga, and Jacobson]{sacht2015nested}
Leonardo Sacht, Etienne Vouga, and Alec Jacobson.
\newblock Nested cages.
\newblock \emph{ACM Transactions on Graphics (TOG)}, 34\penalty0 (6):\penalty0 1--14, 2015.

\bibitem[Saharia et~al.(2022)Saharia, Chan, Saxena, Li, Whang, Denton, Ghasemipour, Gontijo~Lopes, Karagol~Ayan, Salimans, et~al.]{saharia2022photorealistic}
Chitwan Saharia, William Chan, Saurabh Saxena, Lala Li, Jay Whang, Emily~L Denton, Kamyar Ghasemipour, Raphael Gontijo~Lopes, Burcu Karagol~Ayan, Tim Salimans, et~al.
\newblock Photorealistic text-to-image diffusion models with deep language understanding.
\newblock \emph{Advances in neural information processing systems}, 35:\penalty0 36479--36494, 2022.

\bibitem[Sederberg and Parry(1986)]{sederberg1986free}
Thomas~W Sederberg and Scott~R Parry.
\newblock Free-form deformation of solid geometric models.
\newblock In \emph{Proceedings of the 13th annual conference on Computer graphics and interactive techniques}, pages 151--160, 1986.

\bibitem[Shen et~al.(2024)Shen, Xu, Yuan, Yang, Shen, and Wang]{shen2024draggaussian}
Sitian Shen, Jing Xu, Yuheng Yuan, Xingyi Yang, Qiuhong Shen, and Xinchao Wang.
\newblock Draggaussian: Enabling drag-style manipulation on 3d gaussian representation.
\newblock \emph{arXiv preprint arXiv:2405.05800}, 2024.

\bibitem[Shi et~al.(2023)Shi, Wu, Wu, Liu, Zhao, Feng, Liu, Zhang, Zhang, Zhou, et~al.]{shi2023gir}
Yahao Shi, Yanmin Wu, Chenming Wu, Xing Liu, Chen Zhao, Haocheng Feng, Jingtuo Liu, Liangjun Zhang, Jian Zhang, Bin Zhou, et~al.
\newblock Gir: 3d gaussian inverse rendering for relightable scene factorization.
\newblock \emph{arXiv preprint arXiv:2312.05133}, 2023.

\bibitem[Shi et~al.(2024)Shi, Xue, Liew, Pan, Yan, Zhang, Tan, and Bai]{shi2024dragdiffusion}
Yujun Shi, Chuhui Xue, Jun~Hao Liew, Jiachun Pan, Hanshu Yan, Wenqing Zhang, Vincent~YF Tan, and Song Bai.
\newblock Dragdiffusion: Harnessing diffusion models for interactive point-based image editing.
\newblock In \emph{Proceedings of the IEEE/CVF Conference on Computer Vision and Pattern Recognition}, pages 8839--8849, 2024.

\bibitem[Sieger et~al.(2014)Sieger, Menzel, and Botsch]{sieger2014constrained}
Daniel Sieger, Stefan Menzel, and Mario Botsch.
\newblock Constrained space deformation for design optimization.
\newblock \emph{Procedia Engineering}, 82:\penalty0 114--126, 2014.

\bibitem[Sorkine and Alexa(2007)]{sorkine2007rigid}
Olga Sorkine and Marc Alexa.
\newblock As-rigid-as-possible surface modeling.
\newblock In \emph{Symposium on Geometry processing}, pages 109--116. Citeseer, 2007.

\bibitem[Sun et~al.(2024)Sun, Tian, Han, Liu, Zhang, and Xu]{sun2024gseditpro}
Yanhao Sun, Runze Tian, Xiao Han, Xinyao Liu, Yan Zhang, and Kai Xu.
\newblock Gseditpro: 3d gaussian splatting editing with attention-based progressive localization.
\newblock In \emph{Computer Graphics Forum}, page e15215. Wiley Online Library, 2024.

\bibitem[Wand et~al.(2007)Wand, Jenke, Huang, Bokeloh, Guibas, and Schilling]{wand2007reconstruction}
Michael Wand, Philipp Jenke, Qixing Huang, Martin Bokeloh, Leonidas Guibas, and Andreas Schilling.
\newblock Reconstruction of deforming geometry from time-varying point clouds.
\newblock In \emph{Symposium on Geometry processing}, pages 49--58, 2007.

\bibitem[Wang et~al.(2022)Wang, Chai, He, Chen, and Liao]{wang2022clip}
Can Wang, Menglei Chai, Mingming He, Dongdong Chen, and Jing Liao.
\newblock Clip-nerf: Text-and-image driven manipulation of neural radiance fields.
\newblock In \emph{Proceedings of the IEEE/CVF Conference on Computer Vision and Pattern Recognition}, pages 3835--3844, 2022.

\bibitem[Wang et~al.(2024{\natexlab{a}})Wang, Fang, Zhang, Xie, and Tian]{wang2024gaussianeditor}
Junjie Wang, Jiemin Fang, Xiaopeng Zhang, Lingxi Xie, and Qi Tian.
\newblock Gaussianeditor: Editing 3d gaussians delicately with text instructions.
\newblock In \emph{Proceedings of the IEEE/CVF Conference on Computer Vision and Pattern Recognition}, pages 20902--20911, 2024{\natexlab{a}}.

\bibitem[Wang et~al.(2024{\natexlab{b}})Wang, Yue, Zhou, Chan, and Loy]{wang2024exploiting}
Jianyi Wang, Zongsheng Yue, Shangchen Zhou, Kelvin~C.K. Chan, and Chen~Change Loy.
\newblock Exploiting diffusion prior for real-world image super-resolution.
\newblock 2024{\natexlab{b}}.

\bibitem[Wang et~al.(2023)Wang, Saharia, Montgomery, Pont-Tuset, Noy, Pellegrini, Onoe, Laszlo, Fleet, Soricut, et~al.]{wang2023imagen}
Su Wang, Chitwan Saharia, Ceslee Montgomery, Jordi Pont-Tuset, Shai Noy, Stefano Pellegrini, Yasumasa Onoe, Sarah Laszlo, David~J Fleet, Radu Soricut, et~al.
\newblock Imagen editor and editbench: Advancing and evaluating text-guided image inpainting.
\newblock In \emph{Proceedings of the IEEE/CVF conference on computer vision and pattern recognition}, pages 18359--18369, 2023.

\bibitem[Wang et~al.(2025)Wang, Yi, Wu, Zhao, Chen, and Zhang]{wang2025view}
Yuxuan Wang, Xuanyu Yi, Zike Wu, Na Zhao, Long Chen, and Hanwang Zhang.
\newblock View-consistent 3d editing with gaussian splatting.
\newblock In \emph{European Conference on Computer Vision}, pages 404--420. Springer, 2025.

\bibitem[Wu et~al.(2024{\natexlab{a}})Wu, Bian, Li, Wang, Reid, Torr, and Prisacariu]{wu2024gaussctrl}
Jing Wu, Jia-Wang Bian, Xinghui Li, Guangrun Wang, Ian Reid, Philip Torr, and Victor~Adrian Prisacariu.
\newblock Gaussctrl: multi-view consistent text-driven 3d gaussian splatting editing.
\newblock \emph{arXiv preprint arXiv:2403.08733}, 2024{\natexlab{a}}.

\bibitem[Wu et~al.(2024{\natexlab{b}})Wu, Yang, Li, Zhang, Liu, Guibas, Lin, and Wetzstein]{wu2024gpt}
Tong Wu, Guandao Yang, Zhibing Li, Kai Zhang, Ziwei Liu, Leonidas Guibas, Dahua Lin, and Gordon Wetzstein.
\newblock Gpt-4v (ision) is a human-aligned evaluator for text-to-3d generation.
\newblock In \emph{Proceedings of the IEEE/CVF Conference on Computer Vision and Pattern Recognition}, pages 22227--22238, 2024{\natexlab{b}}.

\bibitem[Xu et~al.(2007)Xu, Zhou, Yu, Tan, Peng, and Guo]{xu2007gradient}
Weiwei Xu, Kun Zhou, Yizhou Yu, Qifeng Tan, Qunsheng Peng, and Baining Guo.
\newblock Gradient domain editing of deforming mesh sequences.
\newblock \emph{ACM Transactions On Graphics (TOG)}, 26\penalty0 (3):\penalty0 84--es, 2007.

\bibitem[Ye et~al.(2023)Ye, Danelljan, Yu, and Ke]{ye2023gaussian}
Mingqiao Ye, Martin Danelljan, Fisher Yu, and Lei Ke.
\newblock Gaussian grouping: Segment and edit anything in 3d scenes.
\newblock \emph{arXiv preprint arXiv:2312.00732}, 2023.

\bibitem[Yu et~al.(2024)Yu, Gu, Li, Hu, Kong, Wang, He, Qiao, and Dong]{yu2024scaling}
Fanghua Yu, Jinjin Gu, Zheyuan Li, Jinfan Hu, Xiangtao Kong, Xintao Wang, Jingwen He, Yu Qiao, and Chao Dong.
\newblock Scaling up to excellence: Practicing model scaling for photo-realistic image restoration in the wild.
\newblock In \emph{Proceedings of the IEEE/CVF Conference on Computer Vision and Pattern Recognition}, pages 25669--25680, 2024.

\bibitem[Yu et~al.(2023)Yu, Feng, Feng, Liu, Jin, Zeng, and Chen]{yu2023inpaint}
Tao Yu, Runseng Feng, Ruoyu Feng, Jinming Liu, Xin Jin, Wenjun Zeng, and Zhibo Chen.
\newblock Inpaint anything: Segment anything meets image inpainting.
\newblock \emph{arXiv preprint arXiv:2304.06790}, 2023.

\bibitem[Zhang et~al.(2023)Zhang, Rao, and Agrawala]{zhang2023adding}
Lvmin Zhang, Anyi Rao, and Maneesh Agrawala.
\newblock Adding conditional control to text-to-image diffusion models.
\newblock In \emph{Proceedings of the IEEE/CVF International Conference on Computer Vision}, pages 3836--3847, 2023.

\bibitem[Zhang et~al.(2024)Zhang, Liu, Chen, and Xu]{zhang2024gooddrag}
Zewei Zhang, Huan Liu, Jun Chen, and Xiangyu Xu.
\newblock Gooddrag: Towards good practices for drag editing with diffusion models.
\newblock \emph{arXiv preprint arXiv:2404.07206}, 2024.

\bibitem[Zhuang et~al.(2023)Zhuang, Wang, Lin, Liu, and Li]{zhuang2023dreameditor}
Jingyu Zhuang, Chen Wang, Liang Lin, Lingjie Liu, and Guanbin Li.
\newblock Dreameditor: Text-driven 3d scene editing with neural fields.
\newblock In \emph{SIGGRAPH Asia 2023 Conference Papers}, pages 1--10, 2023.

\bibitem[Zhuang et~al.(2024)Zhuang, Kang, Cao, Li, Lin, and Shan]{zhuang2024tip}
Jingyu Zhuang, Di Kang, Yan-Pei Cao, Guanbin Li, Liang Lin, and Ying Shan.
\newblock Tip-editor: An accurate 3d editor following both text-prompts and image-prompts.
\newblock \emph{ACM Transactions on Graphics (TOG)}, 43\penalty0 (4):\penalty0 1--12, 2024.

\end{thebibliography}
